\pdfoutput=1
\documentclass[aip,
 amsmath,amssymb,
 reprint,%
]{revtex4-2}

\usepackage{graphicx}
\usepackage{dcolumn}
\usepackage{bm}

\usepackage[hidelinks]{hyperref}
\usepackage[utf8]{inputenc}
\usepackage[T1]{fontenc}
\usepackage{mathptmx}
\usepackage{etoolbox}

\makeatletter
\def\@email#1#2{%
 \endgroup
 \patchcmd{\titleblock@produce}
  {\frontmatter@RRAPformat}
  {\frontmatter@RRAPformat{\produce@RRAP{*#1\href{mailto:#2}{#2}}}\frontmatter@RRAPformat}
  {}{}
}%
\makeatother

\usepackage{siunitx}[=v2]
\DeclareUnicodeCharacter{03BB}{\ensuremath{\lambda}}
\DeclareUnicodeCharacter{03C3}{\ensuremath{\sigma}}
\DeclareUnicodeCharacter{2212}{-}
\DeclareSIUnit\gauss{G}

\begin{document}

\preprint{AIP/123-QED}

\title[A grating-chip atomic fountain]{A grating-chip atomic fountain}
\author{Ben Lewis}
\author{Rachel Elvin}
\author{Aidan S.\ Arnold}
\author{Erling Riis}
\author{Paul F.\ Griffin$^*$}
 \email{paul.griffin@strath.ac.uk}
\affiliation{Department of Physics, SUPA, University of Strathclyde, Glasgow G4 0NG, UK}

\date{\today}

\begin{abstract}
Cold atom fountain clocks provide exceptional long term stability as they increase interrogation time at the expense of a larger size. We present a compact cold atom fountain using a grating magneto-optical trap (GMOT) to laser cool and launch the atoms in a simplified optical setup. The fountain is evaluated using coherent population trapping and demonstrates improved single-shot stability from the launch. Ramsey times up to \SI{100}{\ms} were measured with a corresponding fringe linewidth of \SI{5}{\hertz}. This technique could improve both short- and long-term stability of cold atom clocks whilst remaining compact for portable applications.
\end{abstract}

\maketitle

Compact cold-atom systems allow the development of portable, practical, quantum sensors for use outside the laboratory.\cite{takamoto_test_2020,esnault_horace_2011} Cold atoms in free-fall provide a measurement system that is isolated from its environment, allowing high stability and precision.

Atomic systems allow intrinsic accuracy by referencing measurements to the physical constants of the atomic structure.\cite{knappe_microfabricated_2004,kitching_atomic_2011}
The simplest atomic sensors will have a sample of atomic vapour in thermal equilibrium with the environment.
The atoms undergo rapid collisions with each other, the cell walls and any other gas species present.
These interactions cause environmental sensitivity to both pressure and temperature which affect the rate of these collisions.\cite{riley_physics_1992,kozlova_temperature_2011}
To get true accuracy and long-term stability it is necessary to isolate the atoms from their environment as much as possible.
Cold atoms provide this isolation, reducing the collisional interactions of the atoms to negligible levels.

The most accurate microwave clocks -- fountain clocks -- are currently the world's only primary frequency standards.\cite{weyers_advances_2018,szymaniec_npl_2016} These clocks use a Ramsey interrogation sequence which increases the clock accuracy because during the Ramsey time there are virtually no interactions perturbing the atoms.\cite{ramsey_molecular_1950} Launching the atoms upwards in an atomic fountain allows a longer Ramsey time, and increased frequency precision.
This performance comes at the expense of large vacuum chambers, up to \SI{2}{\meter} tall,\cite{ovchinnikov_accurate_2011,hendricks_cs_2019} which are needed to contain the entire parabolic flight path and appropriate magnetic shielding.

\begin{figure}[!b]
    \centering
    \includegraphics{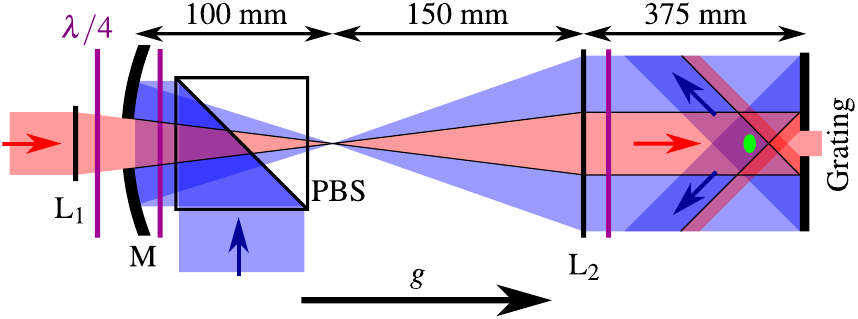}
    \caption{Optics for a GMOT launch (gravity in the $g$ direction, not to scale).  Two concentric beams are produced using a concave mirror with a central hole (M, $f=\SI{100}{\mm}$). The inner beam (red, circular pre-grating profile) hits the atoms (green) only from above, whilst the outer  beam (blue, annular pre-grating profile) hits the atoms only from below. Lens L$_1$ ($f=\SI{200}{\mm}$) brings the inner beam to the same focus as the outer. The mirror surface is imaged onto the grating by lens L$_2$  ($f=\SI{150}{\mm}$) to reduce diffraction effects. Polarising beam-splitter (PBS) and quarter-waveplates ($\lambda/4$) ensure both beams have circular polarisation at the grating.}
    \label{fig:MOTsystem}
\end{figure}

Cold atoms are commonly realised with a 6-beam MOT which requires laser beams incident on the atoms from each Cartesian axis.
This requires bulky optics and good optical access to the atoms. One tool for reducing the volume and complexity of cold-atom trapping is the grating magneto-optical trap (GMOT).\cite{nshii_surface-patterned_2013,mcgilligan_grating_2017} The GMOT uses a diffraction grating to produce overlapping trap beams from a single incident beam, reducing the optical complexity, optical access and volume requirements. GMOTs are  being used to develop compact cold atom systems, sensors and clocks.\cite{bondza_two-color_2022,barker_single-beam_2019,franssen_compact_2019,elvin_towards_2020} 
In this paper, we demonstrate a GMOT-based atomic fountain (Fig.~\ref{fig:MOTsystem}) and characterise its performance using CPT clock signals.


It is possible to launch atoms by pushing with a single resonant laser beam,\cite{kasevich_rf_1989} but this causes significant heating of the atoms via spontaneous emission ($\Delta T \propto \sqrt{N}$, when $N$ photons are scattered) and they will rapidly expand beyond the detection volume. In order to launch \emph{cold} atoms, a `moving molasses' technique is used which maintains the sub-Doppler cooling effect.\cite{clairon_ramsey_1991} This technique requires frequency differences between the MOT beams depending on their orientation relative to the launch direction. The frequency differences are chosen to be cancelled out by the Doppler shift of atoms moving at the launch velocity.

Standard GMOTs have a single input laser beam
with the other beams produced by diffraction. Our $(2\times2)\,$cm$^2$ grating chip comprises three 1D binary grating sectors.\cite{McGilligan2016,Cotter2016} Each sector produces two diffracted orders: one contributing to the MOT and one unused order. All four beams overlap in a central region above the grating giving a `tetrahedral' MOT geometry.\cite{Shimizu1991,Vangeleyn2009} If  all beams originate from a single source, they have the same frequency and form an immobile phase-stable\cite{Hopkins1997} optical lattice (pinned to the grating).

To allow a GMOT-based moving molasses at least two laser frequencies must be present, geometrically separated so as to strike the atoms from different directions. We achieve this using two concentric beams (Fig.~\ref{fig:MOTsystem}). The grating chip has a central hole which prevents the formation of any $0^\mathrm{th}$-order reflection. The hole has a \SI{2}{\mm} diameter as previous work \cite{bregazzi_simple_2021} suggests this is optimal.

The concentric beams are produced (Fig.~\ref{fig:MOTsystem}) using a concave mirror with a central hole.\footnote{a Thorlabs CM254-100CH3-M02} The outer beam is reflected from the mirror, using a polarising beam splitter (PBS) and quarter-wave plate to spatially separate the input and output. The inner beam passes through the mirror aperture, with a second waveplate to match polarisation and a lens (L$_1$) to focus the beam. Both beams form images of the mirror plane near the MOT/grating location. This optical setup produces an inner beam diameter $\approx4.5\,$mm, which should keep diffraction from sharp beam edges well away from the MOT.

To make the total intensity profile of the combined beams as uniform as possible, the inner beam is over-expanded and clips on the mirror aperture. To mitigate diffraction a second lens (L$_2$) is used both to focus a sharp image of the mirror surface onto the grating as well as collimate both beams. The optical power in each beam is individually stabilised to keep an optimal intensity ratio.\footnote{nominally one, but tuned empirically}

The new beam geometry is used to trap and cool $^{87}$Rb atoms in a GMOT with both beams red-detuned from resonance by \SI{9}{\MHz}. A molasses sequence is used: the detuning is increased up to \SI{60}{\MHz} over \SI{2}{\ms}, then the beam intensities are reduced exponentially to 1\% over \SI{1}{\ms}. This yields approximately \num{e7} atoms at  \SI{6}{\micro\kelvin} when unlaunched. To launch the atoms, a frequency difference, $\Delta f$ is imposed between the inner and outer beams for the entire molasses sequence. The required frequency difference is proportional to the desired launch velocity, $v$, and can be calculated as 
\begin{equation}
\label{eqn:speed}
\Delta f = \frac{v}{\lambda}(1+\cos(\theta)),
\end{equation}
where $\lambda=\SI{780}{\nm}$ is the laser wavelength and $\theta = \ang{40}$ is the Bragg diffraction angle off the grating. A \SI{100}{\ms} flight time requires a launch velocity of \SI{0.49}{\meter\per\second} and $\Delta f = \SI{1.1}{\MHz}$.

\begin{figure}[!t]
    \centering
	\includegraphics{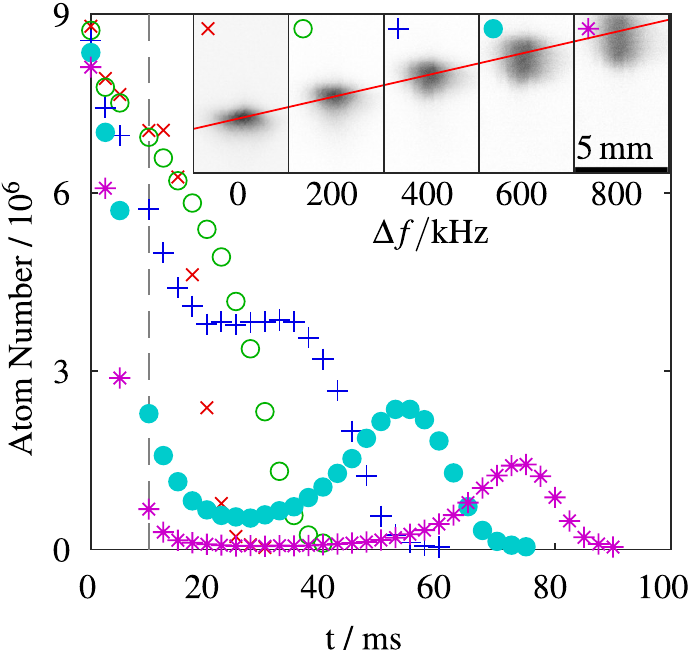}
	\caption{Detected atom number over time for a range of MOT launch beam frequency differences, $\Delta f$. Inset shows a fluorescence image of the MOT at \SI{10}{\milli\second} after launch (dashed line) for each $\Delta f$ value. As $\Delta f$ increases, the MOT is launched faster and it expands more in the launch direction. The red line represents the expected velocity as a function of $\Delta f$ using Eq.~\ref{eqn:speed}.}
	\label{fig:LaunchPerformance}
\end{figure}

Fig.~\ref{fig:LaunchPerformance} shows the performance of the GMOT launch. The atom number at different times after the launch is detected by the absorption from a weak beam perpendicular to the launch.
This beam is aligned with the initial (unlaunched) MOT position, is resonant with the ($F$ = 2 $\to$ 3) transition and has a \SI{7.5}{\mm} $1/e^{2}$ diameter. It is clear from the $\Delta f=$ \SIlist{600;800}{\kHz} data that a launch is occurring because the atoms are observed leaving from and returning to the detection region. However, the number of atoms returning is significantly reduced, largely due to the expansion of the cloud outside of the detection region. Changes to the cloud shape seen in post-launch fluorescence images (Fig.~\ref{fig:LaunchPerformance} inset) suggest that the cooling in the launch direction is less effective at larger launch velocities.

The mechanism for the reduced axial cooling with faster launches is most likely contamination of the moving optical lattice by a `static lattice' component. A `static' lattice component could arise due to light from the inner beam reaching the atoms from below after scattering or diffracting from the edge of the grating hole. Despite this effect we note that our fountain is still competitive with other clocks.  The HORACE clock\cite{esnault_horace_2011} detected \num{e6} atoms after \SI{35}{\ms} and achieved a fractional stability of $\sigma_y(\SI{1}{\second}) = \num{2.2e-13}$, whereas our fountain produces \num{1.5e6} atoms at \SI{75}{\ms}.

For any fountain clock there is a trade-off between size and performance.
A long Ramsey time, $T_\mathrm{R}$, will require a large drift region, as is found in primary Cs fountains. 
The size of drift region required will vary as ${T_\mathrm{R}}^2$, so a significant reduction in size may be achieved without affecting the stability as strongly.
The GMOT launch fits in the region where it provides a significant performance boost over an unlaunched system without requiring significantly more volume; the fountain height of \SI{<13}{\mm} for launch times up to \SI{100}{\ms} could already be in-vacuum for relatively small chambers.
As the fountain size decreases, the short-term stability can also be improved by increasing the cycle rate.
However, there is a limit to this as other elements of the clock cycle can only be shortened so much. 
The loading, launch and detection of the MOT will take at least 10s of \si{\ms} unless a source of precooled atoms is available.\cite{schoser_intense_2002,atutov_fast_2003} Thus $T_\mathrm{R}$ in the range \SIrange{50}{100}{\ms} may allow duty cycles around 50\%, which could allow the interleaving of two systems to suppress the Dick effect.\cite{cheng_suppression_2018}

The impact of the launch on clock operation was evaluated using coherent population trapping (CPT),\cite{elgin_cold-atom_2019} as reported previously.\cite{elvin_cold-atom_2019} In this scheme, the microwave clock frequency is imposed as the beat frequency of a dual-frequency laser beam which puts the atoms into a dark state, phase coherent with the microwave local oscillator (LO). A Ramsey sequence can then measure the difference between LO and atomic frequencies. Phase accumulated between the pulses will cause an increase in the scattering of photons from the CPT beam during the second pulse.

A $\sigma^+$-$\sigma^-$ polarisation \cite{kargapoltsev_high-contrast_2004,liu_high_2017} was used for the CPT probe, as this simplifies the system to one laser and an electro-optic modulator (EOM) operating close to the hyperfine splitting frequency. Both frequency components have the same polarisation, so a polariser can be used to ensure clean polarisations. For each Ramsey sequence, the launch velocity was chosen so that the atoms would be moving upwards and then falling downwards through the CPT beam at the time of the two Ramsey pulses,  respectively. 

The first Ramsey pulse pumps the atoms into a coherent dark state. Ramsey fringes are then seen as an absorption signal in the second pulse. The CPT was retro-reflected through the atoms to give the correct polarisations and prevent Doppler sensitivity. A normalisation scheme is used to cancel intensity fluctuations on the probe beam, as described in \cite{elvin_towards_2020}. A relatively large beam with $1/e^2$ radius of \SI{7.5}{\mm} was used to provide  overlap with the expanded MOT after the fountain. The signal to noise ratio (SNR) of this clock was low (only 10 for a \SI{50}{\ms} Ramsey time) because it measures a few photons of absorption per atom on a background of \SI{20}{\micro\watt}. A similar system using a non-CPT detection scheme would provide greatly increased performance but this system is sufficient for a proof-of-principle.

\begin{figure}[!b]
	\includegraphics{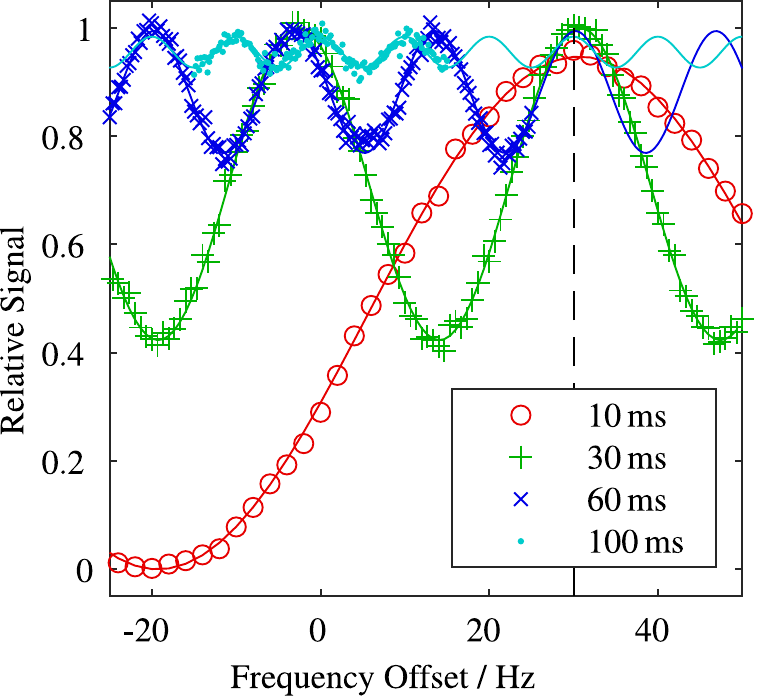}
	\caption{Clock fringes for different Ramsey times. Curves are best-fit sinusoidal fringes to the single-shot data points. The dashed line marks the centre fringe which is shifted \SI{30}{\hertz} from the accepted hyperfine transition frequency due to the second-order Zeeman effect.}
	\label{fig:Fringes}
\end{figure}

Fig.~\ref{fig:Fringes} shows exemplar Ramsey fringes that were observed during this measurement. It is clear that longer Ramsey times (with faster launches) lead to a decrease in both fringe width and signal amplitude. The loss of signal amplitude is consistent with the reduction in atom number due to the launch, as measured by total absorption from the probe beam.
The fringes were measured by varying the frequency of the microwave source around the accepted hyperfine frequency of \SI{6.8346826109}{\GHz}.
The fringes are centred on an offset of $\approx \SI{30}{\Hz}$ which is consistent with the second order Zeeman shift caused by the bias field of \SI{220(10)}{\milli\gauss}.

For each of the Ramsey fringes of Fig.~\ref{fig:Fringes} we can calculate an SNR and fringe width. This allows calculation of the corresponding single-shot fractional stability
\begin{equation}
    \sigma_{y,\mathrm{ss}} = \frac{\sigma_\mathrm{S}}{S_\mathrm{p-p}}\times\frac{1}{2\pi\nu_0T_\mathrm{R}},
    \label{eqn:stability}
\end{equation} with $\sigma_\mathrm{S}$ and $S_\mathrm{p-p}$ being the noise and peak-to-peak amplitude of the signal fringes respectively. The fringe width is determined by $\nu_0$ and $T_\mathrm{R}$, the hyperfine frequency and the Ramsey time respectively.

\begin{figure}[!b]
	\includegraphics{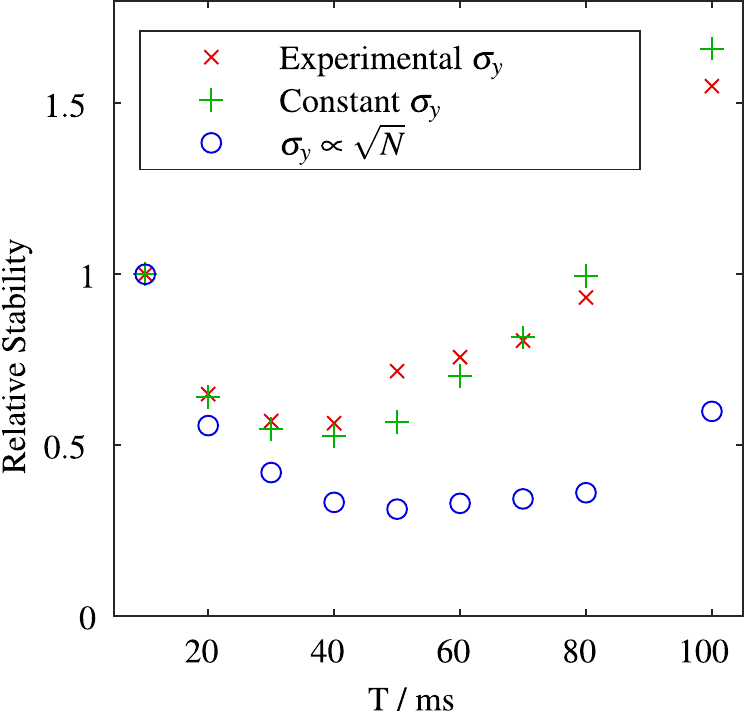}
	\caption{Relative clock stability achieved in a single shot for different Ramsey times. The experimental data (x marks) are a good match to a model with constant noise (+ marks). Quantum projection noise limited measurements (circles) would improve relative performance at longer Ramsey times. For the experimental data, the relative stability of 1 corresponds to an absolute stability of \num{5.5e-11}.}
	\label{fig:Stability}
\end{figure}

The single-shot stability is shown in Fig.~\ref{fig:Stability} and we can also model the stability that we would expect to achieve, using Eq.~\ref{eqn:stability}. We use the experimental fringe amplitude and model the noise amplitude, $\sigma_\mathrm{S}$. Currently, we are limited by electronic and background noise. Therefore, we expect that $\sigma_\mathrm{S}$ should be constant for different $T_\mathrm{R}$. Fig.~\ref{fig:Stability} shows that using a constant value for $\sigma_\mathrm{S}$ gives a good match to the expected stability, except for an outlier at $T_\mathrm{R} = \SI{50}{\ms}$ where $\sigma_\mathrm{S}$ was measured $20\%$ higher than expected. A different clock design with a better detection method could achieve much lower noise and ultimately be limited by the atomic quantum projection noise (QPN).\cite{itano_quantum_1993,santarelli_quantum_1999} In this case we would expect $\sigma_S\propto\sqrt{N}$, for atom number $N$, and the impact of atom loss at longer Ramsey times would be mitigated by decreased noise. Fig.~\ref{fig:Stability} also shows this comparison.

For QPN limited measurements, the best single-shot stability would be observed at $T_\mathrm{R} = \SI{50}{\milli\second}$ but with only a 15\% decrease in stability out to $T_\mathrm{R} = \SI{80}{\milli\second}$. The fountain method would improve stability by a factor of up to 3.3 compared to the \SI{10}{\milli\second} result, which is a reasonable approximation of the performance achievable without a fountain. Additionally, good performance at longer Ramsey times allows the effects of any phase shift induced during the Ramsey pulses to be reduced, improving the accuracy and long-term stability. The consistent performance over a wide range of Ramsey times could be beneficial for an auto-balanced Ramsey scheme.\cite{sanner_autobalanced_2018,yudin_generalized_2018}

We have shown that it is possible to realise a cold atom launch using the compact GMOT. Two concentric beams with a small frequency difference are required. The fountain was evaluated as part of a CPT clock, and found to improve the single-shot stability by increasing the Ramsey time. Over \num{1.5e6} atoms were detected after a time of \SI{75}{\ms}. A best single-shot fractional stability of \num{3.1e-11} was measured. 

The proof-of-principle clock was strongly limited by technical noise and considerable performance improvements should be possible by moving to a lower noise detection scheme. The low optical access requirements will allow integration into compact clock systems.

\begin{acknowledgments}
The authors would like to thank J.~P.\ McGilligan for useful comments on the manuscript. The authors acknowledge Engineering and Physical Sciences Research Council (EP/T001046/1).
\end{acknowledgments}

\section*{Data Availability Statement}
The data that support the findings of
this study are openly available at
https://doi.org/10.15129/3847c7ff-f2f7-473b-8036-bcc4b261c7b8.

\bibliography{references}

\begin{thebibliography}{38}%
\makeatletter
\providecommand \@ifxundefined [1]{%
 \@ifx{#1\undefined}
}%
\providecommand \@ifnum [1]{%
 \ifnum #1\expandafter \@firstoftwo
 \else \expandafter \@secondoftwo
 \fi
}%
\providecommand \@ifx [1]{%
 \ifx #1\expandafter \@firstoftwo
 \else \expandafter \@secondoftwo
 \fi
}%
\providecommand \natexlab [1]{#1}%
\providecommand \enquote  [1]{``#1''}%
\providecommand \bibnamefont  [1]{#1}%
\providecommand \bibfnamefont [1]{#1}%
\providecommand \citenamefont [1]{#1}%
\providecommand \href@noop [0]{\@secondoftwo}%
\providecommand \href [0]{\begingroup \@sanitize@url \@href}%
\providecommand \@href[1]{\@@startlink{#1}\@@href}%
\providecommand \@@href[1]{\endgroup#1\@@endlink}%
\providecommand \@sanitize@url [0]{\catcode `\\12\catcode `\$12\catcode
  `\&12\catcode `\#12\catcode `\^12\catcode `\_12\catcode `\%12\relax}%
\providecommand \@@startlink[1]{}%
\providecommand \@@endlink[0]{}%
\providecommand \url  [0]{\begingroup\@sanitize@url \@url }%
\providecommand \@url [1]{\endgroup\@href {#1}{\urlprefix }}%
\providecommand \urlprefix  [0]{URL }%
\providecommand \Eprint [0]{\href }%
\providecommand \doibase [0]{https://doi.org/}%
\providecommand \selectlanguage [0]{\@gobble}%
\providecommand \bibinfo  [0]{\@secondoftwo}%
\providecommand \bibfield  [0]{\@secondoftwo}%
\providecommand \translation [1]{[#1]}%
\providecommand \BibitemOpen [0]{}%
\providecommand \bibitemStop [0]{}%
\providecommand \bibitemNoStop [0]{.\EOS\space}%
\providecommand \EOS [0]{\spacefactor3000\relax}%
\providecommand \BibitemShut  [1]{\csname bibitem#1\endcsname}%
\let\auto@bib@innerbib\@empty
\bibitem [{\citenamefont {Takamoto}\ \emph {et~al.}(2020)\citenamefont
  {Takamoto}, \citenamefont {Ushijima}, \citenamefont {Ohmae}, \citenamefont
  {Yahagi}, \citenamefont {Kokado}, \citenamefont {Shinkai},\ and\
  \citenamefont {Katori}}]{takamoto_test_2020}%
  \BibitemOpen
  \bibfield  {author} {\bibinfo {author} {\bibfnamefont {M.}~\bibnamefont
  {Takamoto}}, \bibinfo {author} {\bibfnamefont {I.}~\bibnamefont {Ushijima}},
  \bibinfo {author} {\bibfnamefont {N.}~\bibnamefont {Ohmae}}, \bibinfo
  {author} {\bibfnamefont {T.}~\bibnamefont {Yahagi}}, \bibinfo {author}
  {\bibfnamefont {K.}~\bibnamefont {Kokado}}, \bibinfo {author} {\bibfnamefont
  {H.}~\bibnamefont {Shinkai}},\ and\ \bibinfo {author} {\bibfnamefont
  {H.}~\bibnamefont {Katori}},\ }\bibfield  {title} {\enquote {\bibinfo {title}
  {Test of general relativity by a pair of transportable optical lattice
  clocks},}\ }\href {https://doi.org/10.1038/s41566-020-0619-8} {\bibfield
  {journal} {\bibinfo  {journal} {Nat. Photonics}\ }\textbf {\bibinfo {volume}
  {14}},\ \bibinfo {pages} {411--415} (\bibinfo {year} {2020})}\BibitemShut
  {NoStop}%
\bibitem [{\citenamefont {Esnault}\ \emph {et~al.}(2011)\citenamefont
  {Esnault}, \citenamefont {Rossetto}, \citenamefont {Holleville},
  \citenamefont {Delporte},\ and\ \citenamefont
  {Dimarcq}}]{esnault_horace_2011}%
  \BibitemOpen
  \bibfield  {author} {\bibinfo {author} {\bibfnamefont {F.~X.}\ \bibnamefont
  {Esnault}}, \bibinfo {author} {\bibfnamefont {N.}~\bibnamefont {Rossetto}},
  \bibinfo {author} {\bibfnamefont {D.}~\bibnamefont {Holleville}}, \bibinfo
  {author} {\bibfnamefont {J.}~\bibnamefont {Delporte}},\ and\ \bibinfo
  {author} {\bibfnamefont {N.}~\bibnamefont {Dimarcq}},\ }\bibfield  {title}
  {\enquote {\bibinfo {title} {{HORACE}: A compact cold atom clock for
  {G}alileo},}\ }\href {https://doi.org/10.1016/j.asr.2010.12.012} {\bibfield
  {journal} {\bibinfo  {journal} {Advances in Space Research}\ }\textbf
  {\bibinfo {volume} {47}},\ \bibinfo {pages} {854--858} (\bibinfo {year}
  {2011})}\BibitemShut {NoStop}%
\bibitem [{\citenamefont {Knappe}\ \emph {et~al.}(2004)\citenamefont {Knappe},
  \citenamefont {Shah}, \citenamefont {Schwindt}, \citenamefont {Hollberg},
  \citenamefont {Kitching}, \citenamefont {Liew},\ and\ \citenamefont
  {Moreland}}]{knappe_microfabricated_2004}%
  \BibitemOpen
  \bibfield  {author} {\bibinfo {author} {\bibfnamefont {S.}~\bibnamefont
  {Knappe}}, \bibinfo {author} {\bibfnamefont {V.}~\bibnamefont {Shah}},
  \bibinfo {author} {\bibfnamefont {P.~D.~D.}\ \bibnamefont {Schwindt}},
  \bibinfo {author} {\bibfnamefont {L.}~\bibnamefont {Hollberg}}, \bibinfo
  {author} {\bibfnamefont {J.}~\bibnamefont {Kitching}}, \bibinfo {author}
  {\bibfnamefont {L.-A.}\ \bibnamefont {Liew}},\ and\ \bibinfo {author}
  {\bibfnamefont {J.}~\bibnamefont {Moreland}},\ }\bibfield  {title} {\enquote
  {\bibinfo {title} {A microfabricated atomic clock},}\ }\href
  {https://doi.org/10.1063/1.1787942} {\bibfield  {journal} {\bibinfo
  {journal} {Applied Physics Letters}\ }\textbf {\bibinfo {volume} {85}},\
  \bibinfo {pages} {1460--1462} (\bibinfo {year} {2004})}\BibitemShut {NoStop}%
\bibitem [{\citenamefont {Kitching}, \citenamefont {Knappe},\ and\
  \citenamefont {Donley}(2011)}]{kitching_atomic_2011}%
  \BibitemOpen
  \bibfield  {author} {\bibinfo {author} {\bibfnamefont {J.}~\bibnamefont
  {Kitching}}, \bibinfo {author} {\bibfnamefont {S.}~\bibnamefont {Knappe}},\
  and\ \bibinfo {author} {\bibfnamefont {E.~A.}\ \bibnamefont {Donley}},\
  }\bibfield  {title} {\enquote {\bibinfo {title} {Atomic sensors – a
  review},}\ }\href {https://doi.org/10.1109/JSEN.2011.2157679} {\bibfield
  {journal} {\bibinfo  {journal} {{IEEE} Sensors Journal}\ }\textbf {\bibinfo
  {volume} {11}},\ \bibinfo {pages} {1749--1758} (\bibinfo {year}
  {2011})}\BibitemShut {NoStop}%
\bibitem [{\citenamefont {Riley}(1992)}]{riley_physics_1992}%
  \BibitemOpen
  \bibfield  {author} {\bibinfo {author} {\bibfnamefont {W.}~\bibnamefont
  {Riley}},\ }\bibfield  {title} {\enquote {\bibinfo {title} {The physics of
  the environmental sensitivity of rubidium gas cell atomic frequency
  standards},}\ }\href {https://doi.org/10.1109/58.139119} {\bibfield
  {journal} {\bibinfo  {journal} {{IEEE} Transactions on Ultrasonics,
  Ferroelectrics, and Frequency Control}\ }\textbf {\bibinfo {volume} {39}},\
  \bibinfo {pages} {232--240} (\bibinfo {year} {1992})}\BibitemShut {NoStop}%
\bibitem [{\citenamefont {Kozlova}\ \emph {et~al.}(2011)\citenamefont
  {Kozlova}, \citenamefont {Danet}, \citenamefont {Guérandel},\ and\
  \citenamefont {de~Clercq}}]{kozlova_temperature_2011}%
  \BibitemOpen
  \bibfield  {author} {\bibinfo {author} {\bibfnamefont {O.}~\bibnamefont
  {Kozlova}}, \bibinfo {author} {\bibfnamefont {J.-M.}\ \bibnamefont {Danet}},
  \bibinfo {author} {\bibfnamefont {S.}~\bibnamefont {Guérandel}},\ and\
  \bibinfo {author} {\bibfnamefont {E.}~\bibnamefont {de~Clercq}},\ }\bibfield
  {title} {\enquote {\bibinfo {title} {Temperature dependence of a {C}s vapor
  cell clock: Pressure shift, signal amplitude, light shift},}\ }in\ \href
  {https://doi.org/10.1109/FCS.2011.5977839} {\emph {\bibinfo {booktitle} {2011
  Joint Conference of the {IEEE} International Frequency Control and the
  European Frequency and Time Forum ({FCS}) Proceedings}}}\ (\bibinfo {year}
  {2011})\ pp.\ \bibinfo {pages} {1--5}\BibitemShut {NoStop}%
\bibitem [{\citenamefont {Weyers}\ \emph {et~al.}(2018)\citenamefont {Weyers},
  \citenamefont {Gerginov}, \citenamefont {Kazda}, \citenamefont {Rahm},
  \citenamefont {Lipphardt}, \citenamefont {Dobrev},\ and\ \citenamefont
  {Gibble}}]{weyers_advances_2018}%
  \BibitemOpen
  \bibfield  {author} {\bibinfo {author} {\bibfnamefont {S.}~\bibnamefont
  {Weyers}}, \bibinfo {author} {\bibfnamefont {V.}~\bibnamefont {Gerginov}},
  \bibinfo {author} {\bibfnamefont {M.}~\bibnamefont {Kazda}}, \bibinfo
  {author} {\bibfnamefont {J.}~\bibnamefont {Rahm}}, \bibinfo {author}
  {\bibfnamefont {B.}~\bibnamefont {Lipphardt}}, \bibinfo {author}
  {\bibfnamefont {G.}~\bibnamefont {Dobrev}},\ and\ \bibinfo {author}
  {\bibfnamefont {K.}~\bibnamefont {Gibble}},\ }\bibfield  {title} {\enquote
  {\bibinfo {title} {Advances in the accuracy, stability, and reliability of
  the {PTB} primary fountain clocks},}\ }\href
  {https://doi.org/10.1088/1681-7575/aae008} {\bibfield  {journal} {\bibinfo
  {journal} {Metrologia}\ }\textbf {\bibinfo {volume} {55}},\ \bibinfo {pages}
  {789--805} (\bibinfo {year} {2018})}\BibitemShut {NoStop}%
\bibitem [{\citenamefont {Szymaniec}\ \emph {et~al.}(2016)\citenamefont
  {Szymaniec}, \citenamefont {Lea}, \citenamefont {Gibble}, \citenamefont
  {Park}, \citenamefont {Liu},\ and\ \citenamefont
  {Głowacki}}]{szymaniec_npl_2016}%
  \BibitemOpen
  \bibfield  {author} {\bibinfo {author} {\bibfnamefont {K.}~\bibnamefont
  {Szymaniec}}, \bibinfo {author} {\bibfnamefont {S.~N.}\ \bibnamefont {Lea}},
  \bibinfo {author} {\bibfnamefont {K.}~\bibnamefont {Gibble}}, \bibinfo
  {author} {\bibfnamefont {S.~E.}\ \bibnamefont {Park}}, \bibinfo {author}
  {\bibfnamefont {K.}~\bibnamefont {Liu}},\ and\ \bibinfo {author}
  {\bibfnamefont {P.}~\bibnamefont {Głowacki}},\ }\bibfield  {title} {\enquote
  {\bibinfo {title} {{NPL} {C}s fountain frequency standards and the quest for
  the ultimate accuracy},}\ }\href
  {https://doi.org/10.1088/1742-6596/723/1/012003} {\bibfield  {journal}
  {\bibinfo  {journal} {Journal of Physics: Conference Series}\ }\textbf
  {\bibinfo {volume} {723}},\ \bibinfo {pages} {012003} (\bibinfo {year}
  {2016})}\BibitemShut {NoStop}%
\bibitem [{\citenamefont {Ramsey}(1950)}]{ramsey_molecular_1950}%
  \BibitemOpen
  \bibfield  {author} {\bibinfo {author} {\bibfnamefont {N.~F.}\ \bibnamefont
  {Ramsey}},\ }\bibfield  {title} {\enquote {\bibinfo {title} {A molecular beam
  resonance method with separated oscillating fields},}\ }\href
  {https://doi.org/10.1103/PhysRev.78.695} {\bibfield  {journal} {\bibinfo
  {journal} {Physical Review}\ }\textbf {\bibinfo {volume} {78}},\ \bibinfo
  {pages} {695--699} (\bibinfo {year} {1950})}\BibitemShut {NoStop}%
\bibitem [{\citenamefont {Ovchinnikov}\ and\ \citenamefont
  {Marra}(2011)}]{ovchinnikov_accurate_2011}%
  \BibitemOpen
  \bibfield  {author} {\bibinfo {author} {\bibfnamefont {Y.}~\bibnamefont
  {Ovchinnikov}}\ and\ \bibinfo {author} {\bibfnamefont {G.}~\bibnamefont
  {Marra}},\ }\bibfield  {title} {\enquote {\bibinfo {title} {Accurate rubidium
  atomic fountain frequency standard},}\ }\href
  {https://doi.org/10.1088/0026-1394/48/3/003} {\bibfield  {journal} {\bibinfo
  {journal} {Metrologia}\ }\textbf {\bibinfo {volume} {48}},\ \bibinfo {pages}
  {87--100} (\bibinfo {year} {2011})}\BibitemShut {NoStop}%
\bibitem [{\citenamefont {Hendricks}\ \emph {et~al.}(2019)\citenamefont
  {Hendricks}, \citenamefont {Ozimek}, \citenamefont {Szymaniec}, \citenamefont
  {Nagórny}, \citenamefont {Dunst}, \citenamefont {Nawrocki}, \citenamefont
  {Beattie}, \citenamefont {Jian},\ and\ \citenamefont
  {Gibble}}]{hendricks_cs_2019}%
  \BibitemOpen
  \bibfield  {author} {\bibinfo {author} {\bibfnamefont {R.~J.}\ \bibnamefont
  {Hendricks}}, \bibinfo {author} {\bibfnamefont {F.}~\bibnamefont {Ozimek}},
  \bibinfo {author} {\bibfnamefont {K.}~\bibnamefont {Szymaniec}}, \bibinfo
  {author} {\bibfnamefont {B.}~\bibnamefont {Nagórny}}, \bibinfo {author}
  {\bibfnamefont {P.}~\bibnamefont {Dunst}}, \bibinfo {author} {\bibfnamefont
  {J.}~\bibnamefont {Nawrocki}}, \bibinfo {author} {\bibfnamefont
  {S.}~\bibnamefont {Beattie}}, \bibinfo {author} {\bibfnamefont
  {B.}~\bibnamefont {Jian}},\ and\ \bibinfo {author} {\bibfnamefont
  {K.}~\bibnamefont {Gibble}},\ }\bibfield  {title} {\enquote {\bibinfo {title}
  {{C}s fountain clocks for commercial realizations—an improved and robust
  design},}\ }\href {https://doi.org/10.1109/TUFFC.2018.2874550} {\bibfield
  {journal} {\bibinfo  {journal} {{IEEE} Transactions on Ultrasonics,
  Ferroelectrics, and Frequency Control}\ }\textbf {\bibinfo {volume} {66}},\
  \bibinfo {pages} {624--631} (\bibinfo {year} {2019})}\BibitemShut {NoStop}%
\bibitem [{\citenamefont {Nshii}\ \emph {et~al.}(2013)\citenamefont {Nshii},
  \citenamefont {Vangeleyn}, \citenamefont {Cotter}, \citenamefont {Griffin},
  \citenamefont {Hinds}, \citenamefont {Ironside}, \citenamefont {See},
  \citenamefont {Sinclair}, \citenamefont {Riis},\ and\ \citenamefont
  {Arnold}}]{nshii_surface-patterned_2013}%
  \BibitemOpen
  \bibfield  {author} {\bibinfo {author} {\bibfnamefont {C.~C.}\ \bibnamefont
  {Nshii}}, \bibinfo {author} {\bibfnamefont {M.}~\bibnamefont {Vangeleyn}},
  \bibinfo {author} {\bibfnamefont {J.~P.}\ \bibnamefont {Cotter}}, \bibinfo
  {author} {\bibfnamefont {P.~F.}\ \bibnamefont {Griffin}}, \bibinfo {author}
  {\bibfnamefont {E.~A.}\ \bibnamefont {Hinds}}, \bibinfo {author}
  {\bibfnamefont {C.~N.}\ \bibnamefont {Ironside}}, \bibinfo {author}
  {\bibfnamefont {P.}~\bibnamefont {See}}, \bibinfo {author} {\bibfnamefont
  {A.~G.}\ \bibnamefont {Sinclair}}, \bibinfo {author} {\bibfnamefont
  {E.}~\bibnamefont {Riis}},\ and\ \bibinfo {author} {\bibfnamefont {A.~S.}\
  \bibnamefont {Arnold}},\ }\bibfield  {title} {\enquote {\bibinfo {title} {A
  surface-patterned chip as a strong source of ultracold atoms for quantum
  technologies},}\ }\href {https://doi.org/10.1038/nnano.2013.47} {\bibfield
  {journal} {\bibinfo  {journal} {Nature Nanotechnology}\ }\textbf {\bibinfo
  {volume} {8}},\ \bibinfo {pages} {321--324} (\bibinfo {year}
  {2013})}\BibitemShut {NoStop}%
\bibitem [{\citenamefont {{McGilligan}}\ \emph {et~al.}(2017)\citenamefont
  {{McGilligan}}, \citenamefont {Griffin}, \citenamefont {Elvin}, \citenamefont
  {Ingleby}, \citenamefont {Riis},\ and\ \citenamefont
  {Arnold}}]{mcgilligan_grating_2017}%
  \BibitemOpen
  \bibfield  {author} {\bibinfo {author} {\bibfnamefont {J.~P.}\ \bibnamefont
  {{McGilligan}}}, \bibinfo {author} {\bibfnamefont {P.~F.}\ \bibnamefont
  {Griffin}}, \bibinfo {author} {\bibfnamefont {R.}~\bibnamefont {Elvin}},
  \bibinfo {author} {\bibfnamefont {S.~J.}\ \bibnamefont {Ingleby}}, \bibinfo
  {author} {\bibfnamefont {E.}~\bibnamefont {Riis}},\ and\ \bibinfo {author}
  {\bibfnamefont {A.~S.}\ \bibnamefont {Arnold}},\ }\bibfield  {title}
  {\enquote {\bibinfo {title} {Grating chips for quantum technologies},}\
  }\href {https://doi.org/10.1038/s41598-017-00254-0} {\bibfield  {journal}
  {\bibinfo  {journal} {Scientific Reports}\ }\textbf {\bibinfo {volume} {7}},\
  \bibinfo {pages} {384} (\bibinfo {year} {2017})}\BibitemShut {NoStop}%
\bibitem [{\citenamefont {Bondza}\ \emph {et~al.}(2022)\citenamefont {Bondza},
  \citenamefont {Lisdat}, \citenamefont {Kroker},\ and\ \citenamefont
  {Leopold}}]{bondza_two-color_2022}%
  \BibitemOpen
  \bibfield  {author} {\bibinfo {author} {\bibfnamefont {S.}~\bibnamefont
  {Bondza}}, \bibinfo {author} {\bibfnamefont {C.}~\bibnamefont {Lisdat}},
  \bibinfo {author} {\bibfnamefont {S.}~\bibnamefont {Kroker}},\ and\ \bibinfo
  {author} {\bibfnamefont {T.}~\bibnamefont {Leopold}},\ }\bibfield  {title}
  {\enquote {\bibinfo {title} {Two-color grating magneto-optical trap for
  narrow-line laser cooling},}\ }\href
  {https://doi.org/10.1103/PhysRevApplied.17.044002} {\bibfield  {journal}
  {\bibinfo  {journal} {Physical Review Applied}\ }\textbf {\bibinfo {volume}
  {17}},\ \bibinfo {pages} {044002} (\bibinfo {year} {2022})}\BibitemShut
  {NoStop}%
\bibitem [{\citenamefont {Barker}\ \emph {et~al.}(2019)\citenamefont {Barker},
  \citenamefont {Norrgard}, \citenamefont {Klimov}, \citenamefont {Fedchak},
  \citenamefont {Scherschligt},\ and\ \citenamefont
  {Eckel}}]{barker_single-beam_2019}%
  \BibitemOpen
  \bibfield  {author} {\bibinfo {author} {\bibfnamefont {D.}~\bibnamefont
  {Barker}}, \bibinfo {author} {\bibfnamefont {E.}~\bibnamefont {Norrgard}},
  \bibinfo {author} {\bibfnamefont {N.}~\bibnamefont {Klimov}}, \bibinfo
  {author} {\bibfnamefont {J.}~\bibnamefont {Fedchak}}, \bibinfo {author}
  {\bibfnamefont {J.}~\bibnamefont {Scherschligt}},\ and\ \bibinfo {author}
  {\bibfnamefont {S.}~\bibnamefont {Eckel}},\ }\bibfield  {title} {\enquote
  {\bibinfo {title} {Single-beam zeeman slower and magneto-optical trap using a
  nanofabricated grating},}\ }\href
  {https://doi.org/10.1103/PhysRevApplied.11.064023} {\bibfield  {journal}
  {\bibinfo  {journal} {Physical Review Applied}\ }\textbf {\bibinfo {volume}
  {11}},\ \bibinfo {pages} {064023} (\bibinfo {year} {2019})}\BibitemShut
  {NoStop}%
\bibitem [{\citenamefont {Franssen}\ \emph {et~al.}(2019)\citenamefont
  {Franssen}, \citenamefont {de~Raadt}, \citenamefont {van Ninhuijs},\ and\
  \citenamefont {Luiten}}]{franssen_compact_2019}%
  \BibitemOpen
  \bibfield  {author} {\bibinfo {author} {\bibfnamefont {J.}~\bibnamefont
  {Franssen}}, \bibinfo {author} {\bibfnamefont {T.}~\bibnamefont {de~Raadt}},
  \bibinfo {author} {\bibfnamefont {M.}~\bibnamefont {van Ninhuijs}},\ and\
  \bibinfo {author} {\bibfnamefont {O.}~\bibnamefont {Luiten}},\ }\bibfield
  {title} {\enquote {\bibinfo {title} {Compact ultracold electron source based
  on a grating magneto-optical trap},}\ }\href
  {https://doi.org/10.1103/PhysRevAccelBeams.22.023401} {\bibfield  {journal}
  {\bibinfo  {journal} {Physical Review Accelerators and Beams}\ }\textbf
  {\bibinfo {volume} {22}},\ \bibinfo {pages} {023401} (\bibinfo {year}
  {2019})}\BibitemShut {NoStop}%
\bibitem [{\citenamefont {Elvin}\ \emph {et~al.}(2020)\citenamefont {Elvin},
  \citenamefont {Wright}, \citenamefont {Lewis}, \citenamefont {Keliehor},
  \citenamefont {Bregazzi}, \citenamefont {{McGilligan}}, \citenamefont
  {Arnold}, \citenamefont {Griffin},\ and\ \citenamefont
  {Riis}}]{elvin_towards_2020}%
  \BibitemOpen
  \bibfield  {author} {\bibinfo {author} {\bibfnamefont {R.}~\bibnamefont
  {Elvin}}, \bibinfo {author} {\bibfnamefont {M.~W.}\ \bibnamefont {Wright}},
  \bibinfo {author} {\bibfnamefont {B.}~\bibnamefont {Lewis}}, \bibinfo
  {author} {\bibfnamefont {B.~L.}\ \bibnamefont {Keliehor}}, \bibinfo {author}
  {\bibfnamefont {A.}~\bibnamefont {Bregazzi}}, \bibinfo {author}
  {\bibfnamefont {J.~P.}\ \bibnamefont {{McGilligan}}}, \bibinfo {author}
  {\bibfnamefont {A.~S.}\ \bibnamefont {Arnold}}, \bibinfo {author}
  {\bibfnamefont {P.~F.}\ \bibnamefont {Griffin}},\ and\ \bibinfo {author}
  {\bibfnamefont {E.}~\bibnamefont {Riis}},\ }\bibfield  {title} {\enquote
  {\bibinfo {title} {Towards a compact, optically interrogated, cold-atom
  microwave clock},}\ }\href {https://doi.org/10.1515/aot-2020-0022} {\bibfield
   {journal} {\bibinfo  {journal} {Advanced Optical Technologies}\ }\textbf
  {\bibinfo {volume} {9}},\ \bibinfo {pages} {297--303} (\bibinfo {year}
  {2020})}\BibitemShut {NoStop}%
\bibitem [{\citenamefont {Kasevich}\ \emph {et~al.}(1989)\citenamefont
  {Kasevich}, \citenamefont {Riis}, \citenamefont {Chu},\ and\ \citenamefont
  {{DeVoe}}}]{kasevich_rf_1989}%
  \BibitemOpen
  \bibfield  {author} {\bibinfo {author} {\bibfnamefont {M.~A.}\ \bibnamefont
  {Kasevich}}, \bibinfo {author} {\bibfnamefont {E.}~\bibnamefont {Riis}},
  \bibinfo {author} {\bibfnamefont {S.}~\bibnamefont {Chu}},\ and\ \bibinfo
  {author} {\bibfnamefont {R.~G.}\ \bibnamefont {{DeVoe}}},\ }\bibfield
  {title} {\enquote {\bibinfo {title} {{RF} spectroscopy in an atomic
  fountain},}\ }\href {https://doi.org/10.1103/PhysRevLett.63.612} {\bibfield
  {journal} {\bibinfo  {journal} {Physical Review Letters}\ }\textbf {\bibinfo
  {volume} {63}},\ \bibinfo {pages} {612--615} (\bibinfo {year}
  {1989})}\BibitemShut {NoStop}%
\bibitem [{\citenamefont {Clairon}\ \emph {et~al.}(1991)\citenamefont
  {Clairon}, \citenamefont {Salomon}, \citenamefont {Guellati},\ and\
  \citenamefont {Phillips}}]{clairon_ramsey_1991}%
  \BibitemOpen
  \bibfield  {author} {\bibinfo {author} {\bibfnamefont {A.}~\bibnamefont
  {Clairon}}, \bibinfo {author} {\bibfnamefont {C.}~\bibnamefont {Salomon}},
  \bibinfo {author} {\bibfnamefont {S.}~\bibnamefont {Guellati}},\ and\
  \bibinfo {author} {\bibfnamefont {W.~D.}\ \bibnamefont {Phillips}},\
  }\bibfield  {title} {\enquote {\bibinfo {title} {Ramsey resonance in a
  zacharias fountain},}\ }\href {https://doi.org/10.1209/0295-5075/16/2/008}
  {\bibfield  {journal} {\bibinfo  {journal} {Europhysics Letters ({EPL})}\
  }\textbf {\bibinfo {volume} {16}},\ \bibinfo {pages} {165--170} (\bibinfo
  {year} {1991})}\BibitemShut {NoStop}%
\bibitem [{\citenamefont {McGilligan}\ \emph {et~al.}(2016)\citenamefont
  {McGilligan}, \citenamefont {Griffin}, \citenamefont {Riis},\ and\
  \citenamefont {Arnold}}]{McGilligan2016}%
  \BibitemOpen
  \bibfield  {author} {\bibinfo {author} {\bibfnamefont {J.~P.}\ \bibnamefont
  {McGilligan}}, \bibinfo {author} {\bibfnamefont {P.~F.}\ \bibnamefont
  {Griffin}}, \bibinfo {author} {\bibfnamefont {E.}~\bibnamefont {Riis}},\ and\
  \bibinfo {author} {\bibfnamefont {A.~S.}\ \bibnamefont {Arnold}},\ }\bibfield
   {title} {\enquote {\bibinfo {title} {Diffraction-grating characterization
  for cold-atom experiments},}\ }\href
  {https://doi.org/10.1364/josab.33.001271} {\bibfield  {journal} {\bibinfo
  {journal} {Journal of the Optical Society of America B}\ }\textbf {\bibinfo
  {volume} {33}},\ \bibinfo {pages} {1271} (\bibinfo {year}
  {2016})}\BibitemShut {NoStop}%
\bibitem [{\citenamefont {Cotter}\ \emph {et~al.}(2016)\citenamefont {Cotter},
  \citenamefont {McGilligan}, \citenamefont {Griffin}, \citenamefont {Rabey},
  \citenamefont {Docherty}, \citenamefont {Riis}, \citenamefont {Arnold},\ and\
  \citenamefont {Hinds}}]{Cotter2016}%
  \BibitemOpen
  \bibfield  {author} {\bibinfo {author} {\bibfnamefont {J.~P.}\ \bibnamefont
  {Cotter}}, \bibinfo {author} {\bibfnamefont {J.~P.}\ \bibnamefont
  {McGilligan}}, \bibinfo {author} {\bibfnamefont {P.~F.}\ \bibnamefont
  {Griffin}}, \bibinfo {author} {\bibfnamefont {I.~M.}\ \bibnamefont {Rabey}},
  \bibinfo {author} {\bibfnamefont {K.}~\bibnamefont {Docherty}}, \bibinfo
  {author} {\bibfnamefont {E.}~\bibnamefont {Riis}}, \bibinfo {author}
  {\bibfnamefont {A.~S.}\ \bibnamefont {Arnold}},\ and\ \bibinfo {author}
  {\bibfnamefont {E.~A.}\ \bibnamefont {Hinds}},\ }\bibfield  {title} {\enquote
  {\bibinfo {title} {Design and fabrication of diffractive atom chips for laser
  cooling and trapping},}\ }\href {https://doi.org/10.1007/s00340-016-6415-y}
  {\bibfield  {journal} {\bibinfo  {journal} {Applied Physics B}\ }\textbf
  {\bibinfo {volume} {122}},\ \bibinfo {pages} {172} (\bibinfo {year}
  {2016})}\BibitemShut {NoStop}%
\bibitem [{\citenamefont {Shimizu}, \citenamefont {Shimizu},\ and\
  \citenamefont {Takuma}(1991)}]{Shimizu1991}%
  \BibitemOpen
  \bibfield  {author} {\bibinfo {author} {\bibfnamefont {F.}~\bibnamefont
  {Shimizu}}, \bibinfo {author} {\bibfnamefont {K.}~\bibnamefont {Shimizu}},\
  and\ \bibinfo {author} {\bibfnamefont {H.}~\bibnamefont {Takuma}},\
  }\bibfield  {title} {\enquote {\bibinfo {title} {Four-beam laser trap of
  neutral atoms},}\ }\href {https://doi.org/10.1364/ol.16.000339} {\bibfield
  {journal} {\bibinfo  {journal} {Optics Letters}\ }\textbf {\bibinfo {volume}
  {16}},\ \bibinfo {pages} {339} (\bibinfo {year} {1991})}\BibitemShut
  {NoStop}%
\bibitem [{\citenamefont {Vangeleyn}\ \emph {et~al.}(2009)\citenamefont
  {Vangeleyn}, \citenamefont {Griffin}, \citenamefont {Riis},\ and\
  \citenamefont {Arnold}}]{Vangeleyn2009}%
  \BibitemOpen
  \bibfield  {author} {\bibinfo {author} {\bibfnamefont {M.}~\bibnamefont
  {Vangeleyn}}, \bibinfo {author} {\bibfnamefont {P.~F.}\ \bibnamefont
  {Griffin}}, \bibinfo {author} {\bibfnamefont {E.}~\bibnamefont {Riis}},\ and\
  \bibinfo {author} {\bibfnamefont {A.~S.}\ \bibnamefont {Arnold}},\ }\bibfield
   {title} {\enquote {\bibinfo {title} {Single-laser, one beam, tetrahedral
  magneto-optical trap},}\ }\href {https://doi.org/10.1364/oe.17.013601}
  {\bibfield  {journal} {\bibinfo  {journal} {Optics Express}\ }\textbf
  {\bibinfo {volume} {17}},\ \bibinfo {pages} {13601} (\bibinfo {year}
  {2009})}\BibitemShut {NoStop}%
\bibitem [{\citenamefont {Hopkins}\ and\ \citenamefont
  {Durrant}(1997)}]{Hopkins1997}%
  \BibitemOpen
  \bibfield  {author} {\bibinfo {author} {\bibfnamefont {S.~A.}\ \bibnamefont
  {Hopkins}}\ and\ \bibinfo {author} {\bibfnamefont {A.~V.}\ \bibnamefont
  {Durrant}},\ }\bibfield  {title} {\enquote {\bibinfo {title} {Parameters for
  polarization gradients in three-dimensional electromagnetic standing
  waves},}\ }\href {https://doi.org/10.1103/physreva.56.4012} {\bibfield
  {journal} {\bibinfo  {journal} {Physical Review A}\ }\textbf {\bibinfo
  {volume} {56}},\ \bibinfo {pages} {4012--4022} (\bibinfo {year}
  {1997})}\BibitemShut {NoStop}%
\bibitem [{\citenamefont {Bregazzi}\ \emph {et~al.}(2021)\citenamefont
  {Bregazzi}, \citenamefont {Griffin}, \citenamefont {Arnold}, \citenamefont
  {Burt}, \citenamefont {Martinez}, \citenamefont {Boudot}, \citenamefont
  {Kitching}, \citenamefont {Riis},\ and\ \citenamefont
  {{McGilligan}}}]{bregazzi_simple_2021}%
  \BibitemOpen
  \bibfield  {author} {\bibinfo {author} {\bibfnamefont {A.}~\bibnamefont
  {Bregazzi}}, \bibinfo {author} {\bibfnamefont {P.~F.}\ \bibnamefont
  {Griffin}}, \bibinfo {author} {\bibfnamefont {A.~S.}\ \bibnamefont {Arnold}},
  \bibinfo {author} {\bibfnamefont {D.~P.}\ \bibnamefont {Burt}}, \bibinfo
  {author} {\bibfnamefont {G.}~\bibnamefont {Martinez}}, \bibinfo {author}
  {\bibfnamefont {R.}~\bibnamefont {Boudot}}, \bibinfo {author} {\bibfnamefont
  {J.}~\bibnamefont {Kitching}}, \bibinfo {author} {\bibfnamefont
  {E.}~\bibnamefont {Riis}},\ and\ \bibinfo {author} {\bibfnamefont {J.~P.}\
  \bibnamefont {{McGilligan}}},\ }\bibfield  {title} {\enquote {\bibinfo
  {title} {A simple imaging solution for chip-scale laser cooling},}\ }\href
  {https://doi.org/10.1063/5.0068725} {\bibfield  {journal} {\bibinfo
  {journal} {Applied Physics Letters}\ }\textbf {\bibinfo {volume} {119}},\
  \bibinfo {pages} {184002} (\bibinfo {year} {2021})}\BibitemShut {NoStop}%
\bibitem [{Note1()}]{Note1}%
  \BibitemOpen
  \bibinfo {note} {A Thorlabs CM254-100CH3-M02}\BibitemShut {NoStop}%
\bibitem [{Note2()}]{Note2}%
  \BibitemOpen
  \bibinfo {note} {Nominally one, but tuned empirically}\BibitemShut {NoStop}%
\bibitem [{\citenamefont {Schoser}\ \emph {et~al.}(2002)\citenamefont
  {Schoser}, \citenamefont {Batär}, \citenamefont {Löw}, \citenamefont
  {Schweikhard}, \citenamefont {Grabowski}, \citenamefont {Ovchinnikov},\ and\
  \citenamefont {Pfau}}]{schoser_intense_2002}%
  \BibitemOpen
  \bibfield  {author} {\bibinfo {author} {\bibfnamefont {J.}~\bibnamefont
  {Schoser}}, \bibinfo {author} {\bibfnamefont {A.}~\bibnamefont {Batär}},
  \bibinfo {author} {\bibfnamefont {R.}~\bibnamefont {Löw}}, \bibinfo {author}
  {\bibfnamefont {V.}~\bibnamefont {Schweikhard}}, \bibinfo {author}
  {\bibfnamefont {A.}~\bibnamefont {Grabowski}}, \bibinfo {author}
  {\bibfnamefont {Y.~B.}\ \bibnamefont {Ovchinnikov}},\ and\ \bibinfo {author}
  {\bibfnamefont {T.}~\bibnamefont {Pfau}},\ }\bibfield  {title} {\enquote
  {\bibinfo {title} {Intense source of cold {R}b atoms from a pure
  two-dimensional magneto-optical trap},}\ }\href
  {https://doi.org/10.1103/PhysRevA.66.023410} {\bibfield  {journal} {\bibinfo
  {journal} {Physical Review A}\ }\textbf {\bibinfo {volume} {66}},\ \bibinfo
  {pages} {023410} (\bibinfo {year} {2002})}\BibitemShut {NoStop}%
\bibitem [{\citenamefont {Atutov}\ \emph {et~al.}(2003)\citenamefont {Atutov},
  \citenamefont {Calabrese}, \citenamefont {Guidi}, \citenamefont {Mai},
  \citenamefont {Rudavets}, \citenamefont {Scansani}, \citenamefont
  {Tomassetti}, \citenamefont {Biancalana}, \citenamefont {Burchianti},
  \citenamefont {Marinelli}, \citenamefont {Mariotti}, \citenamefont {Moi},\
  and\ \citenamefont {Veronesi}}]{atutov_fast_2003}%
  \BibitemOpen
  \bibfield  {author} {\bibinfo {author} {\bibfnamefont {S.~N.}\ \bibnamefont
  {Atutov}}, \bibinfo {author} {\bibfnamefont {R.}~\bibnamefont {Calabrese}},
  \bibinfo {author} {\bibfnamefont {V.}~\bibnamefont {Guidi}}, \bibinfo
  {author} {\bibfnamefont {B.}~\bibnamefont {Mai}}, \bibinfo {author}
  {\bibfnamefont {A.~G.}\ \bibnamefont {Rudavets}}, \bibinfo {author}
  {\bibfnamefont {E.}~\bibnamefont {Scansani}}, \bibinfo {author}
  {\bibfnamefont {L.}~\bibnamefont {Tomassetti}}, \bibinfo {author}
  {\bibfnamefont {V.}~\bibnamefont {Biancalana}}, \bibinfo {author}
  {\bibfnamefont {A.}~\bibnamefont {Burchianti}}, \bibinfo {author}
  {\bibfnamefont {C.}~\bibnamefont {Marinelli}}, \bibinfo {author}
  {\bibfnamefont {E.}~\bibnamefont {Mariotti}}, \bibinfo {author}
  {\bibfnamefont {L.}~\bibnamefont {Moi}},\ and\ \bibinfo {author}
  {\bibfnamefont {S.}~\bibnamefont {Veronesi}},\ }\bibfield  {title} {\enquote
  {\bibinfo {title} {Fast and efficient loading of a {R}b magneto-optical trap
  using light-induced atomic desorption},}\ }\href
  {https://doi.org/10.1103/PhysRevA.67.053401} {\bibfield  {journal} {\bibinfo
  {journal} {Physical Review A}\ }\textbf {\bibinfo {volume} {67}},\ \bibinfo
  {pages} {053401} (\bibinfo {year} {2003})}\BibitemShut {NoStop}%
\bibitem [{\citenamefont {Cheng}\ \emph {et~al.}(2018)\citenamefont {Cheng},
  \citenamefont {Sun}, \citenamefont {Zhang},\ and\ \citenamefont
  {Wang}}]{cheng_suppression_2018}%
  \BibitemOpen
  \bibfield  {author} {\bibinfo {author} {\bibfnamefont {P.}~\bibnamefont
  {Cheng}}, \bibinfo {author} {\bibfnamefont {X.}~\bibnamefont {Sun}}, \bibinfo
  {author} {\bibfnamefont {J.}~\bibnamefont {Zhang}},\ and\ \bibinfo {author}
  {\bibfnamefont {L.}~\bibnamefont {Wang}},\ }\bibfield  {title} {\enquote
  {\bibinfo {title} {Suppression of {D}ick effect in ramsey-{CPT} atomic clock
  by interleaving lock},}\ }\href {https://doi.org/10.1109/TUFFC.2018.2864622}
  {\bibfield  {journal} {\bibinfo  {journal} {{IEEE} Transactions on
  Ultrasonics, Ferroelectrics, and Frequency Control}\ }\textbf {\bibinfo
  {volume} {65}},\ \bibinfo {pages} {2195--2200} (\bibinfo {year}
  {2018})}\BibitemShut {NoStop}%
\bibitem [{\citenamefont {Elgin}\ \emph {et~al.}(2019)\citenamefont {Elgin},
  \citenamefont {Heavner}, \citenamefont {Kitching}, \citenamefont {Donley},
  \citenamefont {Denney},\ and\ \citenamefont {Salim}}]{elgin_cold-atom_2019}%
  \BibitemOpen
  \bibfield  {author} {\bibinfo {author} {\bibfnamefont {J.~D.}\ \bibnamefont
  {Elgin}}, \bibinfo {author} {\bibfnamefont {T.~P.}\ \bibnamefont {Heavner}},
  \bibinfo {author} {\bibfnamefont {J.}~\bibnamefont {Kitching}}, \bibinfo
  {author} {\bibfnamefont {E.~A.}\ \bibnamefont {Donley}}, \bibinfo {author}
  {\bibfnamefont {J.}~\bibnamefont {Denney}},\ and\ \bibinfo {author}
  {\bibfnamefont {E.~A.}\ \bibnamefont {Salim}},\ }\bibfield  {title} {\enquote
  {\bibinfo {title} {A cold-atom beam clock based on coherent population
  trapping},}\ }\href {https://doi.org/10.1063/1.5087119} {\bibfield  {journal}
  {\bibinfo  {journal} {Applied Physics Letters}\ }\textbf {\bibinfo {volume}
  {115}},\ \bibinfo {pages} {033503} (\bibinfo {year} {2019})}\BibitemShut
  {NoStop}%
\bibitem [{\citenamefont {Elvin}\ \emph {et~al.}(2019)\citenamefont {Elvin},
  \citenamefont {Hoth}, \citenamefont {Wright}, \citenamefont {Lewis},
  \citenamefont {{McGilligan}}, \citenamefont {{McGilligan}}, \citenamefont
  {Arnold}, \citenamefont {Griffin},\ and\ \citenamefont
  {Riis}}]{elvin_cold-atom_2019}%
  \BibitemOpen
  \bibfield  {author} {\bibinfo {author} {\bibfnamefont {R.}~\bibnamefont
  {Elvin}}, \bibinfo {author} {\bibfnamefont {G.~W.}\ \bibnamefont {Hoth}},
  \bibinfo {author} {\bibfnamefont {M.}~\bibnamefont {Wright}}, \bibinfo
  {author} {\bibfnamefont {B.}~\bibnamefont {Lewis}}, \bibinfo {author}
  {\bibfnamefont {J.~P.}\ \bibnamefont {{McGilligan}}}, \bibinfo {author}
  {\bibfnamefont {J.~P.}\ \bibnamefont {{McGilligan}}}, \bibinfo {author}
  {\bibfnamefont {A.~S.}\ \bibnamefont {Arnold}}, \bibinfo {author}
  {\bibfnamefont {P.~F.}\ \bibnamefont {Griffin}},\ and\ \bibinfo {author}
  {\bibfnamefont {E.}~\bibnamefont {Riis}},\ }\bibfield  {title} {\enquote
  {\bibinfo {title} {Cold-atom clock based on a diffractive optic},}\ }\href
  {https://doi.org/10.1364/OE.378632} {\bibfield  {journal} {\bibinfo
  {journal} {Optics Express}\ }\textbf {\bibinfo {volume} {27}},\ \bibinfo
  {pages} {38359--38366} (\bibinfo {year} {2019})}\BibitemShut {NoStop}%
\bibitem [{\citenamefont {Kargapoltsev}\ \emph {et~al.}(2004)\citenamefont
  {Kargapoltsev}, \citenamefont {Kitching}, \citenamefont {Hollberg},
  \citenamefont {Taichenachev}, \citenamefont {Velichansky},\ and\
  \citenamefont {Yudin}}]{kargapoltsev_high-contrast_2004}%
  \BibitemOpen
  \bibfield  {author} {\bibinfo {author} {\bibfnamefont {S.~V.}\ \bibnamefont
  {Kargapoltsev}}, \bibinfo {author} {\bibfnamefont {J.}~\bibnamefont
  {Kitching}}, \bibinfo {author} {\bibfnamefont {L.}~\bibnamefont {Hollberg}},
  \bibinfo {author} {\bibfnamefont {A.~V.}\ \bibnamefont {Taichenachev}},
  \bibinfo {author} {\bibfnamefont {V.~L.}\ \bibnamefont {Velichansky}},\ and\
  \bibinfo {author} {\bibfnamefont {V.~I.}\ \bibnamefont {Yudin}},\ }\bibfield
  {title} {\enquote {\bibinfo {title} {High-contrast dark resonance in σ+ -
  σ- optical field},}\ }\href {https://doi.org/10.1002/lapl.200410107}
  {\bibfield  {journal} {\bibinfo  {journal} {Laser Physics Letters}\ }\textbf
  {\bibinfo {volume} {1}},\ \bibinfo {pages} {495} (\bibinfo {year}
  {2004})}\BibitemShut {NoStop}%
\bibitem [{\citenamefont {Liu}\ \emph {et~al.}(2017)\citenamefont {Liu},
  \citenamefont {Yudin}, \citenamefont {Taichenachev}, \citenamefont
  {Kitching},\ and\ \citenamefont {Donley}}]{liu_high_2017}%
  \BibitemOpen
  \bibfield  {author} {\bibinfo {author} {\bibfnamefont {X.}~\bibnamefont
  {Liu}}, \bibinfo {author} {\bibfnamefont {V.~I.}\ \bibnamefont {Yudin}},
  \bibinfo {author} {\bibfnamefont {A.~V.}\ \bibnamefont {Taichenachev}},
  \bibinfo {author} {\bibfnamefont {J.}~\bibnamefont {Kitching}},\ and\
  \bibinfo {author} {\bibfnamefont {E.~A.}\ \bibnamefont {Donley}},\ }\bibfield
   {title} {\enquote {\bibinfo {title} {High contrast dark resonances in a
  cold-atom clock probed with counterpropagating circularly polarized beams},}\
  }\href {https://doi.org/10.1063/1.5001179} {\bibfield  {journal} {\bibinfo
  {journal} {Applied Physics Letters}\ }\textbf {\bibinfo {volume} {111}},\
  \bibinfo {pages} {224102} (\bibinfo {year} {2017})}\BibitemShut {NoStop}%
\bibitem [{\citenamefont {Itano}\ \emph {et~al.}(1993)\citenamefont {Itano},
  \citenamefont {Bergquist}, \citenamefont {Bollinger}, \citenamefont
  {Gilligan}, \citenamefont {Heinzen}, \citenamefont {Moore}, \citenamefont
  {Raizen},\ and\ \citenamefont {Wineland}}]{itano_quantum_1993}%
  \BibitemOpen
  \bibfield  {author} {\bibinfo {author} {\bibfnamefont {W.~M.}\ \bibnamefont
  {Itano}}, \bibinfo {author} {\bibfnamefont {J.~C.}\ \bibnamefont
  {Bergquist}}, \bibinfo {author} {\bibfnamefont {J.~J.}\ \bibnamefont
  {Bollinger}}, \bibinfo {author} {\bibfnamefont {J.~M.}\ \bibnamefont
  {Gilligan}}, \bibinfo {author} {\bibfnamefont {D.~J.}\ \bibnamefont
  {Heinzen}}, \bibinfo {author} {\bibfnamefont {F.~L.}\ \bibnamefont {Moore}},
  \bibinfo {author} {\bibfnamefont {M.~G.}\ \bibnamefont {Raizen}},\ and\
  \bibinfo {author} {\bibfnamefont {D.~J.}\ \bibnamefont {Wineland}},\
  }\bibfield  {title} {\enquote {\bibinfo {title} {Quantum projection noise:
  Population fluctuations in two-level systems},}\ }\href
  {https://doi.org/10.1103/PhysRevA.47.3554} {\bibfield  {journal} {\bibinfo
  {journal} {Physical Review A}\ }\textbf {\bibinfo {volume} {47}},\ \bibinfo
  {pages} {3554--3570} (\bibinfo {year} {1993})}\BibitemShut {NoStop}%
\bibitem [{\citenamefont {Santarelli}\ \emph {et~al.}(1999)\citenamefont
  {Santarelli}, \citenamefont {Laurent}, \citenamefont {Lemonde}, \citenamefont
  {Clairon}, \citenamefont {Mann}, \citenamefont {Chang}, \citenamefont
  {Luiten},\ and\ \citenamefont {Salomon}}]{santarelli_quantum_1999}%
  \BibitemOpen
  \bibfield  {author} {\bibinfo {author} {\bibfnamefont {G.}~\bibnamefont
  {Santarelli}}, \bibinfo {author} {\bibfnamefont {P.}~\bibnamefont {Laurent}},
  \bibinfo {author} {\bibfnamefont {P.}~\bibnamefont {Lemonde}}, \bibinfo
  {author} {\bibfnamefont {A.}~\bibnamefont {Clairon}}, \bibinfo {author}
  {\bibfnamefont {A.~G.}\ \bibnamefont {Mann}}, \bibinfo {author}
  {\bibfnamefont {S.}~\bibnamefont {Chang}}, \bibinfo {author} {\bibfnamefont
  {A.~N.}\ \bibnamefont {Luiten}},\ and\ \bibinfo {author} {\bibfnamefont
  {C.}~\bibnamefont {Salomon}},\ }\bibfield  {title} {\enquote {\bibinfo
  {title} {Quantum projection noise in an atomic fountain: A high stability
  cesium frequency standard},}\ }\href
  {https://doi.org/10.1103/PhysRevLett.82.4619} {\bibfield  {journal} {\bibinfo
   {journal} {Physical Review Letters}\ }\textbf {\bibinfo {volume} {82}},\
  \bibinfo {pages} {4619--4622} (\bibinfo {year} {1999})}\BibitemShut {NoStop}%
\bibitem [{\citenamefont {Sanner}\ \emph {et~al.}(2018)\citenamefont {Sanner},
  \citenamefont {Huntemann}, \citenamefont {Lange}, \citenamefont {Tamm},\ and\
  \citenamefont {Peik}}]{sanner_autobalanced_2018}%
  \BibitemOpen
  \bibfield  {author} {\bibinfo {author} {\bibfnamefont {C.}~\bibnamefont
  {Sanner}}, \bibinfo {author} {\bibfnamefont {N.}~\bibnamefont {Huntemann}},
  \bibinfo {author} {\bibfnamefont {R.}~\bibnamefont {Lange}}, \bibinfo
  {author} {\bibfnamefont {C.}~\bibnamefont {Tamm}},\ and\ \bibinfo {author}
  {\bibfnamefont {E.}~\bibnamefont {Peik}},\ }\bibfield  {title} {\enquote
  {\bibinfo {title} {Autobalanced ramsey spectroscopy},}\ }\href
  {https://doi.org/10.1103/PhysRevLett.120.053602} {\bibfield  {journal}
  {\bibinfo  {journal} {Physical Review Letters}\ }\textbf {\bibinfo {volume}
  {120}},\ \bibinfo {pages} {053602} (\bibinfo {year} {2018})}\BibitemShut
  {NoStop}%
\bibitem [{\citenamefont {Yudin}\ \emph {et~al.}(2018)\citenamefont {Yudin},
  \citenamefont {Taichenachev}, \citenamefont {Basalaev}, \citenamefont
  {Zanon-Willette}, \citenamefont {Pollock}, \citenamefont {Shuker},
  \citenamefont {Donley},\ and\ \citenamefont
  {Kitching}}]{yudin_generalized_2018}%
  \BibitemOpen
  \bibfield  {author} {\bibinfo {author} {\bibfnamefont {V.}~\bibnamefont
  {Yudin}}, \bibinfo {author} {\bibfnamefont {A.}~\bibnamefont {Taichenachev}},
  \bibinfo {author} {\bibfnamefont {M.~Y.}\ \bibnamefont {Basalaev}}, \bibinfo
  {author} {\bibfnamefont {T.}~\bibnamefont {Zanon-Willette}}, \bibinfo
  {author} {\bibfnamefont {J.}~\bibnamefont {Pollock}}, \bibinfo {author}
  {\bibfnamefont {M.}~\bibnamefont {Shuker}}, \bibinfo {author} {\bibfnamefont
  {E.}~\bibnamefont {Donley}},\ and\ \bibinfo {author} {\bibfnamefont
  {J.}~\bibnamefont {Kitching}},\ }\bibfield  {title} {\enquote {\bibinfo
  {title} {Generalized autobalanced ramsey spectroscopy of clock
  transitions},}\ }\href {https://doi.org/10.1103/PhysRevApplied.9.054034}
  {\bibfield  {journal} {\bibinfo  {journal} {Physical Review Applied}\
  }\textbf {\bibinfo {volume} {9}},\ \bibinfo {pages} {054034} (\bibinfo {year}
  {2018})}\BibitemShut {NoStop}%
\end{thebibliography}%

\end{document}